\documentstyle[11pt,aasms4,psfig]{article}

\def\asca{{\it ASCA}}

\def\axaf{{\it AXAF}}

\def\egret{{\it EGRET}}
\def\einstein{{\it Einstein}}

\def\rosat{{\it ROSAT}}
\def\xmm{{\it XMM}}

\def\mydegree{^\circ\mskip-5mu}
\def\myarcmin{^\prime\mskip-5mu}
\def\myarcsec{^{\prime\prime}\mskip-5mu}

\begin{document}

\title{X-ray Observation and Analysis of The Composite \\
Supernova Remnant G327.1-1.1}

\author{Ming Sun\altaffilmark{2,3}, Zhen-ru Wang\altaffilmark{1,2} and Yang Chen\altaffilmark{2,4}
\altaffiltext{1}{Center of Astronomy and Astrophysics, CCAST(World Laboratory), zrwang@nju.edu.cn}
\altaffiltext{2}{Department of Astronomy, Nanjing University, Nanjing, 210093, P.R.China}
\altaffiltext{3}{quqiny@nju.edu.cn}
\altaffiltext{4}{ygchen@nju.edu.cn}}

\begin{abstract}
\par\indent
Based on the data from the observation of the supernova remnant(SNR) G327.1-1.1
by {\it Advanced Satellite for Cosmology and Astrophysics} (\asca) and \rosat\ , we find that G327.1-1.1 is a composite remnant with both a nonthermal emission component
and a diffuse thermal emission component. The nonthermal component is well fitted by a power-law model with 
photon index $\Gamma \sim$ 2.2. This component is attributed to the emission from the synchrotron nebula powered by an undiscovered central pulsar. 
The thermal component has a temperature of 
about 0.4 keV. We attribute it to the emission from the shock-heat swept-up ISM. 
Its age, explosion energy and density of ambient medium are derived from the observed thermal component.
 Some charactistics about the synchrotron nebula are also derived. We search for the pulsed signal, but 
has not found it. 
The soft X-ray(0.4 - 2 keV) and hard X-ray(2 - 10 keV) images are different, but they both elongate 
in the SE-NW direction. And this X-ray SE-NW elongation is in positional coincidence with the radio
ridge in MOST 843MHz radio map. 
We present a possibility that the X-ray nonthermal emission mainly come from the trail
produced by a quickly moving undiscoverd pulsar, and the long radio ridge
is formed when the pulsar is moving out of the boundary of the plerionic structure.

\end{abstract}


\section{INTRODUCTION}

Supernova Remnants(SNRs) are usually classified as shell-type, plerion-type and composite-type(Weiler
1985). Composite-type is the most complex class with various appearance. These different
appearance in X-ray and radio morphology and emission nature are owing to different characteristics
of their progenitors and different physical conditions of interstellar medium(ISM) surrounding them.
 For composite SNRs it is important to investigate
the emission nature of the center-brightened part. 
 As to the nonthermal nature, it is usually believed to come from the synchrotron
nebula powered by pulsar wind(e.g. , Blanton \& Helfand 1996; Tamura et al.1996; Harrus, Hughes \& Slane 1998). In recent years, 
the \asca\ (Tanaka, Inoue \& Holt 1994) observations have revealed more and more nonthermal nature in the central part of 
some SNRs, such as G11.2$-$0.3(Vasisht et al.1996), Kes75(Blanton \& Helfand 1996), W44(Harrus, Hughes \& Helfand 1996), MSH15-52(Tamura et al.1996), CTA1(Slane et al.1997), 
Kes73(Gotthelf \& Vasisht 1997), MSH11-62(Harrus et al.1998). 
These findings will improve our knowledge in the evolution of massive stars, the evolution of the synchrotron nebula
 and the occuring rates of two kinds of SNe.

G327.1-1.1 was detected by Caswell et al.(1975) as a faint shell with a diameter about 14$\myarcmin\mbox{ }$ in radio band. Lamb and Markert(1981) found that it was located in the error box of 
the Cos-B source CG327-0. 
Seward (1990), in his summary work on \einstein\ observation of SNRs, classified it as an 
"irregular" SNR of which image has somewhat elongation in south and east-south direction.
 Based on the \rosat\ observation, Seward, Kearns \& Rhode(1996) used a simple blast-wave model to estimate its age 
at about 7000years, a low ISM density at about 0.065 electrons\mbox{ }cm$^{-3}$ and the supernova
explosion energy at about 0.13$\times10^{51}$ergs under an assumed distance of 6.5kpc.
 Whiteoak and Green(1996) found that this source was a composite SNR in 843MHz with a faint shell
and an unusual off-centre plerionic component(Fig.4), and the plerionic component has two peaks 
in the central
part. They also noticed that a 2$\myarcmin\mbox{  }$ long ridge extended north-west from the
west side of the plerionic component. The contribution of the plerionic component is 2.0 Jy, while the total flux of G327.1-1.1 in 843MHz is 7.6 Jy. 
 No public radio spectral index of G327.1-1.1 is available. Since its flux in 408MHz is 10.6 Jy and 
flux in 5GHz is 4.3 Jy(Clark, Caswell \& Green 1975), as an estimate,  its radio spectral index for the whole 
SNR is about 0.4.
 
In this paper, we present the \asca\ observation of G327.1-1.1 and the analyses combined with \rosat\ 
observation and MOST 843MHz radio observation. \asca\ \& \rosat\ observation and data reduction are in 
$\S$2. The 
analyses of its spectra, spatial structure and the temporal behavior are in $\S$3. The derived physical
quantities and physical scenario are discussed in $\S$4 . $\S$5 is a simple conclusion.
 
\section{X-RAY OBSERVATION AND DATA REDUCTION}

\asca\ observed G327.1-1.1 on March 9, 1996 with two kinds of detectors, the gas scintillation 
spectrometer(GIS) and 
the solid state spectrometer(SIS). SIS profits from higher energy resolution($E/{\Delta}E$=50, at 6 keV) than GIS, but
has poor time resolution(4s for 1-CCD mode) and its spatial resolution is limited by the wide point-spread 
function(PSF) of x-ray telescope(XRT) with FWHM of about 3$\myarcmin$. GIS has higher time 
resolution and detection effiency above $\sim$ 3keV than
SIS, so the two kinds of detectors are complementary. In this observation the SIS data is in 1-CCD mode. All data are screened using the 
standard REV2 processing,  
which includes removing photons with low cut-off rigidity(COR), correction for South Atlantic Anomaly (SAA), grade selection, 
removing hot and flickering pixels and so on. After screening we obtain about 22000 events from each GIS detector, about
6000 events from SIS1 and about 7500 events from SIS0. Approximate exposure time is 37.1ks for SIS screened data
and 39.5ks for GIS screened data. 
GIS data used non-standard mode(10 bits for time record) to increase time resolution. To archieve this high
time resolution, the GIS rise-time information(RTI) were sacrificed. Without RTI, the background rejection based on the RTI can not be
done(The \asca\  Data Reduction Guide, 1997), so the data will have more internal background than otherwise.
 Thus
in the spectral analysis, only SIS and PSPC data are used, while GIS data are used to do 
temporal analysis. The standard software FTOOL4.0(XSELECT1.3) are used to process the data.

\rosat\ observed G327.1-1.1 in 1993 March 3-13 using PSPC({\it Position Sensitive Proportional Counter}). After subtracting the background source and foreground source, a total about 1600 events(including background) are obtained in the region of G327.1-1.1. The \rosat\ PSPC data have better spatial 
resolution, larger field of view(FOV) than the \asca\ data, so we use these profits  
as a supplement to the \asca\ data.

\section{ANALYSIS}

\subsection{Spectral analysis}
 
Since G327.1-1.1 is a faint source,
 it is necessary to get a good estimate to its background spectrum.
 The \asca\  Data Reduction Guide provides two basic methods
to obtain background spectra. One is to subtract a source-free part from the same
observation, and the other is to subtract suitable portions of blank-sky
 observations. In the case of G327.1-1.1, the first is not suitable because the observation was
made in 1-CCD mode and the FOV is almost full of the source. In addition,
 the responses of SIS depend on position in the FOV, so background events should
be extracted from the same part of the FOV as the source
events. The second method is good for high latitude sources. But for sources in galactic plane,like 
G327.1-1.1, the contribution of "Galactic ridge" emission(Warwick et al.1985; Koyama et al.1986;
Valinia \& marshall,1998) to the observed spectra must be considered(Blanton \& Helfand 1996).
So the following method is applied to get the background spectrum. First, 
we select a small source-free region in the same observation, and estimate the photon
flux in this region. We also find some observations from the \asca\ archive data that are
near G327.1-1.1 and in each of those observations we also select
one or two small source-free regions and estimate the photon fluxs. If the X-ray background does not
change significantly around G327.1-1.1, those photon fluxs that are estimated in the above regions 
give the approximate range of the background photon flux around G327.1-1.1. Second, we consider
the sequences of the public Galactic-Ridge-data. The photon flux is estimated for each of them.
Those sequences that have photon fluxs approximately in the range that we estimate above 
are selected. Finally the selected sequences are used to extract background spectrum.
 
Refering to the article of
\rosat\ observation on G327.1-1.1(Seward et al.1996) it is found that the source 4(Seward et al.1996
, Fig.5 \& Table 3) is just in the FOV of SIS. 
Although it is weak in the FOV of SIS, we subtract it in the analysis. 
 
After doing this we extract G327.1-1.1 spectra with a 4$\myarcmin$.5 radius circle centered
in 15$^{h}$54$^{m}$16$^{s}$ , -55$\mydegree\mbox{ }$03$\myarcmin\mbox{ }$57$\myarcsec$(J2000). 
The spectra are binned to 
256 channels to contain enough photon counts for $\chi^{2}$ model fits. 
 The inclusion of PSPC data would contribute
to constrain the N$_{\rm H}$. The Raymond \& Smith (1977) thermal plasma model with cosmic abundances
 is used for thermal
component and the power-law model is used for nonthermal component, and the model from Morrison \& McCammon (1983) for
absorption along the line of sight with an equivalent column density of hydrogen is also used. 
The standard software XSPEC(10.0) is used to fit the spectra. 
The result of spectral fit is in Table 1. We can see that adding a thermal component significantly
improves the fit of only a power law model. The spectrum of G327.1-1.1 and the model are shown in Fig.1 
and Fig.2 respectively. It is obviouly that the power-law component is dominant and the thermal component
is distinct only below 2 keV. 
The residuals in Fig.1 show a weak sign for a emission line in 1.3keV - 1.4keV,
 which may be the Mg K$\alpha$ emission line or the blend of Fe-L lines. 

Limited by the FOV of SIS, the region that we study in the above is mainly the central part.
 Therefore, \rosat\
PSPC data are used to analyze the outer part of G327.1-1.1. The data are extracted from a circle of radius 7.5$\myarcmin$ , deducted the part that we have studied in the above and
the foreground and background sources. When performing spectral fitting $N_{\rm H}$ is fixed to $1.8\times10^{22}$cm$^{-2}$,
 according to our above SIS \& PSPC data joint fitting. The fitting result is listed in Table 2.

\subsection{Spatial analysis}

The PSF of the \asca\ X-ray telescopes (XRT) has a relatively sharp core (FWHM $\sim$ $50\myarcsec$ ) but broad wings(half-power 
diameter of 3$\myarcmin$ ). So it is not trivial to do spatial analysis for extended sources.
Here we use the "Lucy-Richardson" method to do deconvolution of SIS images(Jalota, Gotthelf \& Zoonematkermani 1993). 
First the raw soft 
X-ray(0.4 keV-2 keV) image and hard X-ray(2 keV-10 keV) image of G327.1-1.1 are got. Then we
perform background-subtraction, exposure-correction and vignetting-correction on these raw images.
In those processes, FTOOL ASCAEFFMAP and ASCAEXPO are used and the background images are obtained using
a similar method as that in $\S$3.1. Second, we consider applying the "Lucy-Richardson" method.
The signal-to-noise ratio of the two images are all larger than the lower limit of 5$\sigma$.
 But the photon numbers of each image(about 3200 counts) are 
not good enough. So we perform only 18 iterations of the "Lucy-Richardson" 
deconvolution to these two images and smooth the deconvoluted images with a Gaussian function of FWHM=45$\myarcsec$  to avoid the false over-resolving.
To PSPC image, due to its better spatial resolution than \asca\ , we only smooth it with a Gaussian function of FWHM=30$\myarcsec$.  These images are shown in Fig.3.
 The peak of X-ray emission in the soft, hard \asca\ images and the PSPC images are all located 
in the same place: RA = $15^{h}54^{m}25^{s}$($\pm$01$^{s}$) , DEC = -55$\mydegree$ 04$\myarcmin$ 04$\myarcsec$($\pm$09$\myarcsec$)(J2000). And a similar SE-NW elongation structure can be seen in all of the
three images, especially in the hard X-ray image.
 But comparing to the
hard \asca\ image in which the emission mainly come from the places close to the common peak, the soft 
\asca\ image and the PSPC image are more extended. From the best fit the thermal flux is 
estimated at 30$\%$ of total observed flux in soft X-ray band(0.4 keV-2 keV), and we can see some
clumps in PSPC image, which are beyond the hard contours but within the soft contours. Thus,
it may be natural to attribute the soft extension to the contribution of the thermal component.
From $\S$3.1 it is known that almost all the emission above 2keV is nonthermal.
Therefore, the SE-NW elongation in hard X-ray image indicates the SE-NW elongation of the sychrotron nebula. 
 When we overlay the \rosat\ image on the MOST 843MHz radio contour(Fig.4), it is surprising to find that
the X-ray elongation is in positional coincidence with the radio long ridge very well.
 This phenomenon helps to confirm the reality of the SE-NW elongation. If we prolong the
elongation(or radio ridge) to the inner part of the plerion, it will not pass through the brightest
central part. But we notice that there are some weak emission to the SW of the plerion(Fig.4), and if
these are also taken as parts of the plerion, the elongation will roughly in the diameter
of the plerion, so we suggest that an 
undiscovered pulsar(see $\S$4.3) may be located just in the head of the radio ridge
and the X-ray elongation, and the X-ray emission may mainly come from the trail caused by the quick movement of the assumed pulsar, such as Wang, Li \& Begelman (1993), Frail et al.(1996), Wang \& Gotthelf(1998).
This is just a possible interpretation.

\subsection{Temporal analysis}
 
Since the X-ray emission of G327.1-1.1 mainly come from the power-law component, it is interesting to
know if a pulsed signal could be found in
\asca\ data. After barycentered the photon times of arrival(using FTOOL TIMECONV), We extract GIS2(GIS3) light curves from a 5$\myarcmin$ radius circle centered on the peak of the X-ray
emission. Both medium bit-rate mode(resolution $\sim$ 0.5ms) and high bit-rate mode(resolution $\sim$ 0.06ms) data are used. The combination of high and
midium bit-rate mode limits the overall accuracy to 5ms. 
First we set the time bins as 1 min, 2 min and 10 min, no evidences for significant variations on these timescales are found.
 Then we extract only high-energy photons( 2keV - 10keV ) and use several time bins: 5ms , 16ms , 0.1s , 1s. A $2^{20}$ point
fast Fourier transform was applied to the light curves using diferent NEWBIN to search a period down to 10ms, but no 
significant pulsations are detected.
   
\section{DISCUSSION}

\subsection{Distance}
The following methods are used to estimate the distance of G327.1-1.1. 
Ryter , Cesarsky \& Audouze (1975) obtained the following statistics relation : $N_{\rm x}/E(B-V)=(6.8\pm1.6)\times
10^{21}$ equivalent H atoms cm$^{-2}$mag$^{-1}$, where $N_{\rm x}$ refers to equivalent H column 
density observed in X-ray band. From the contour diagrams given by Lucke(1978), 
it is found that $E(B-V)/d \sim$ 0.3mag$\mbox{ }$kpc$^{-1}$ in the direction of G327.1-1.1. 
Combining these two relation and the fitting result of $N_{\rm x}$, we get that the distance of
G327.1-1.1 is about 9kpc.

It is known that SNR MSH15-56 is only about 1$\mydegree$.5 from G327.1-1.1, and the kinematic distance of MSH15-56 is 4.1$\pm$0.7 kpc
(Rasado et al.1996). From the \rosat\ data analysis,Kassim , Hertz \& Weiler (1993) obtained $N_{\rm H}=8.9\pm0.3\times10^{21}$cm$^{-2}$ for MSH15-56.
 Since the measured X-ray absorbing column density of G327.1-1.1 is about twice that measured
for MSH15-56, it is reasonable to assign a distance about 8 - 9 kpc to G327.1-1.1.

Therefore, a distance of 9kpc is used in the following discussion.

\subsection{The thermal component}
As stated in the $\S$3.1, the region that we do the joint spectral fit of the SIS and PSPC data
is mainly the inner part, while the PSPC data are also used to study the other parts
of G327.1-1.1. As shown in Table 1 and 2 , the thermal temperaures obtained for the two regions 
are very similar and the difference between their thermal fluxs can be well explained by the 
difference between their emission volume. Such consistence suggest that the thermal component is 
almost all over the source, and should be diffuse as implied by few condensation in the outer part
of the image and its low flux nature.
The large radius of radio shell of G327.1-1.1 implies its old age. Therefore, it is natural to
think that the SNR is in the adiabatic expansion
phase. For simplicity and
following Seward et al.(1996), we also assume a complete shell, and attribute the incompleteness of the shell to the inhomogenities in the
ISM. A simple Sedov(1959) relation is as the following.
\begin{equation}
\upsilon_{s} = 9.2\times10^{2} (T/keV)^{1/2} (0.6/\mu)^{1/2}
\end{equation}
\begin{equation}
t_{4} = (T/keV)^{-1/2}(R_{s}/25pc) 
\end{equation}
\begin{equation}
E_{51} = 1.16\times10^{-3}(n_{0}/1cm^{-3})(T/keV)(R_{s}/pc)^{3}
\end{equation}
\begin{equation}
M_{su}(M_{\odot}) \simeq 0.10(n_{0}/1cm^{-3})(R_{s}/pc)^{3}
\end{equation}

Where $\upsilon_{\rm s}$ is the shock speed in units of kms$^{-1}$, $t_{4}$ is the time since the supernova explosion in units of 10$^{4}$yr, $n_{0}$ is the mean ambient particle density, 
$E_{51}$ is the explosion energy in units of $10^{51}$ergs, $R_{\rm s}$ is the radius of supernova shock front, $M_{\rm su}$ is the swept-up mass
in units of the mass of sun, $\mu$ = 0.6 for cosmic abundances.

The instantaneous power radiated from a shell is $L_{\rm s}(t) = (16\pi/3)R_{rm s}^{3}(t)n_{0}^{2}\Lambda(T)$, where $\Lambda(T)$ = 1.0$\times
10^{-22}T_{6}^{-0.7}+2.3\times10^{-24}T_{6}^{0.5}$ergs$\mbox{ }$cm$^{3}\mbox{ }$s$^{-1}$ is the radiative cooling function(McCray 1987). Here $T_{6}$
is the postshock temperature in units of $10^{6}\mbox{ }$K. 
We synthesize the fitting results from \asca\ data and \rosat\ data to calculate
the properties and list the results in Table 2. The value of M$_{su}$ is greater than any
reasonable estimate of the initial ejected mass and serves as a self-consistency check to ensure
that G327.1-1.1 has reached the adiabatic phase as assumed. 
The explosion energy and the mean ambient particle density are a little low and
these results are consistent with the result of Seward et al.(1996). Since $E_{51} \propto F_{\rm s}^{1/2}d^{5/2}T/\Lambda^{1/2}$, where $F_{s}$ = $L_{\rm s}$ / 4$\pi$$d^{2}$, 
if assuming that $F_{s} , T$ and $\Lambda $ are relatively fixed , $E_{51}$ will be sensitive to d. If we take a distance value of
 14kpc, $E_{51}$ will be about 0.7, a more canonical value. 
The derived total unabsorded thermal flux is estimated as about 9$\times10^{-11}$ergs\mbox{ }cm$^{-2}$s$^{-1}$.

\subsection{The nonthermal component}
The presence of a dominant power-law component in the \asca\ spectra of G327.1-1.1 clearly
suggests the existence of a undiscovered pulsar and a pulsar-driven synchrontron nebula.
 Based on the observed quantities, some properties of the pulsar and its surrounding synchrontron 
nebula are discussed in follows. 

The luminosity of the nonthermal component and the current spin-down energy loss rate of the pulsar
were found to follow an empirical relationship(Seward \& Wang 1988) for the \einstein\ energy band.
A similar relationship for the \asca\ band(1 - 10keV) is $\log L_{x}$ = 1.27 $\log \dot{E}$ - 12.3 (Kawai, Tamura \& Shibata 1997). 
If taking the observed unabsorbed value of $L_{x}$ we obtain $\dot{E}$ = 1.5$\times10^{37}d^{1.6}_{9}$ergs$\mbox{ }$s$^{-1}$.
 Under the assumption that a pulsar's current spin period is much larger than its initial spin 
period and its moment of inertia, $I = 10^{45}$g cm$^{2}$, we will get the current spin period(Seward \& Wang 1988),
\begin{equation}
P = 0.25[t(10^{3} {\rm yr})]^{-1/2}[\dot{E}(10^{37} {\rm ergs\mbox{ }s^{-1}})]^{-1/2}{\rm s} = 62d^{-1.3}_{9} {\rm ms}
\end{equation}

the period derivative,
\begin{equation}
\dot{P} = 1.58\times10^{-11}[P({\rm s})][t(10^{3} {\rm yr})]^{-1} {\rm s\mbox{ }s^{-1}} = 8.9 \times 10^{-14} d^{-2.3}_{9} {\rm s\mbox{ }s^{-1}}
\end{equation}

the surface magnetic field,
\begin{equation}
B_{0} = [P({\rm s})\dot{P}(10^{-15}\mbox{ }{\rm s\mbox{ }s^{-1}})]^{1/2}\times10^{12}\mbox{ }{\rm G} = 2.3 \times 10^{12} d^{-1.8}_{9} {\rm G}
\end{equation}

Although the radio spectral index of the plerion is unknown, the break frequency $\nu_{B}$
can also be estimated in the two extreme cases. 
In case I, the plerion has a flat spectrum in radio band, $S_{\rm \nu}$ $\sim$ 2Jy. 
Since $S_{\rm \nu}$ $\approx$ 0.3 $\mu$Jy at 5keV is known from the above X-ray spectral fit, 
it is easy to use the derived X-ray spectral index to extrapolate and obtain $\nu_{\rm B}$ $\sim$
2.5$\times10^{3}$ GHz. In case II, assuming a radio spectral index of about 0.4
(as mentioned in $\S$1 for the whole source), then the derived $\nu_{\rm B}$ would be about 
1.3$\times10^{5}$
GHz; or assuming the steepest spectral index 0.3 for pure plerions(Weiler \& Sramek 1988), 
then $\nu_{\rm B}$ would be about 3.5$\times10^{4}$ GHz. 
Hence, $\nu_{\rm B}$ may be in the infrared range: 2.5$\times10^{3}$ GHz - 
1.3$\times10^{5}$ GHz(or 3.5$\times10^{4}$ GHz), which is comparable with the break frequency of the Crab nebula $\sim$ $10^{4}$ GHz, and the derived break frequency of Kes75 $\sim$
$10^{4}$ GHz (Blanton \& Helfand 1996). In the two extreme cases, integrating the spectrum from $10^{7}$ Hz to $10^{11}$ Hz, we derive the radio luminosity
$L_{\rm R}$ of $1.9\times10^{34} $d$^{2}_{9} $ergs$\mbox{ }$s$^{-1} , 4.8\times10^{33} $d$^{2}_{9} $ergs$\mbox{ }$s$^{-1}$ (or 6.6$\times10^{33} $d$^{2}_{9} $ergs$\mbox{ }$s$^{-1}$) respectively, while as a comparison the unabsorbed X-ray luminosity
$L_{\rm X}$ is $1.2\times10^{35} $d$^{2}_{9} $ergs$\mbox{ }$s$^{-1}$. If taking a $\nu_{\rm B}$ value of $10^{4}$ GHz, from the relationship $\nu_{\rm B} \approx 3.4B^{-3}(t/10^{4}$yr$)^{-2}$(Lozinskaya, 1992),  
we get $B \sim$ 0.7$\times10^{-4}$ G. If assuming that the reverse shock has passed the boundary of the plerion, and the equilibrium between the 
plerion magnetic pressure and the ambient thermal pressure $P_{0}$ has been approached , 
from the equilibrium relation: $P_{0}=B^{2}/8\pi$, we can obtain another estimate about $B$ . According to
the Sedov solution, the central pressure should be: $P_{0}=0.31P$(postshock) = 0.0372$\rho_{0}(R_{\rm s}/t)^{2}$ ds (Reynolds \& Chevalier,1984), then we obtained  $P_{0}$=1.4$\times10^{-10}$dyne
cm$^{-2},B \sim$ 0.6$\times10^{-4}$ G. These two estimated values of field strength are similar and
are typical for synchrontron nebulae,such as W44(Frail et al.\ 1996).

The X-ray emission life-time of relativistic particle $\tau_{\rm X}$ is about $50\varepsilon^{-1/2}B_{-4}^{-3/2}$yr, while the radio emission life-time of relativistic particle $\tau_{\rm R}$ is about
$9\times10^{5}B_{-4}^{-3/2}(\nu/1$GHz)$^{-1/2}$yr, where $\varepsilon$ is the characteristic 
synchrontron photon energy in units of keV,and $B_{-4} = B/(10^{-4}G) $. 
Because $\tau_{\rm R} \gg \tau_{\rm X}$, the X-ray synchrotron nebula should be 
significantly smaller than the radio one and should be located close to the pulsar. Therefore, the 
undiscovered pulsar should be located close to the hard X-ray emission.

\subsection{The X-ray trail-like configuration and the morphology of the plerion}
An interesting phenomenon of this source is the unusual X-ray trail-like morphology that we have mentioned above. 
As a possible explanation, we attribute this trail-like morphology to the trail confined by a bow 
shock which is caused by a quickly moving undiscovered pulsar.
The X-ray emission might mainly come from the shocked pulsar wind, which flows gradually from the head 
portion of the bow shock to the trail along the thick layer of the reverse shock. Assuming
the pulsar was "kicked" out at the moment of supernova explosion and the birthplace is the
geometric center of the plerion, the projectional displacement of the pulsar is about 7pc and the
derived velocity of the pulsar would be, $\upsilon_{\rm p}$ $\sim$ 600(sin$\theta$)$^{-1}$km $s^{-1}$,
where $\theta$ is the angle between the line of sight and the real direction of the trail. 
This derived velocity is a reasonable   
velocity for pulsars(Lyne \& Lorimer 1994). The ram pressure balance condition, n$_{0}$m$_{\rm H}\upsilon_{p}^{2}=\mbox{ }\dot{E}/(4\pi r_{\rm a}^{2}c)$, yields the standoff distance of the apex of the bow shock
, $r_{\rm a}$ $\sim$ 0.082($n_{0}$/0.1cm$^{-3})^{-1/2}$sin$\theta$ pc. This distance corresponds
to the observed separation of about 1$\myarcsec$.9 and obviously can not be resolved by \asca\ or \rosat\ . 
The situation here seems like that in PSR1929+10(Wang et al.\ 1993), where the X-ray radiation
arise from relativistic flows inside a narrow tunnel. For estimating the length of the trail,
a quantitative analysis yields a bulk velocity $\sim$ 0.5 - 0.6c 
in the layer of the shocked pulsar wind in most portion of the tunnel, based on the algorithm 
given by Chen, Bandiera \& Wang (1996). With such a bulk velocity, in the X-ray emitting lifetime,
 the X-ray synchrotron-emitting flow (typified by 1keV) will go through a length of about 7.5 - 9pc.
This is essentially in agreement with the messured projected length about 6pc when considering
the factor, sin$\theta$ and the uncertainty in the value of $B$.
 
G327.1-1.1 is clearly a composite SNR in the radio band. The plerion is nearly in the center
of the SNR, and the little eastward displacement may be explained by the inhomogenities in the ISM. 
The ridge may be created by the drag of the quickly moving pulsar. 
Certainly further observations are necessary to obtain more information about this source, such as
the spatial distribution of the radio spectral index, the information about polarization, etc. 
 
If a quickly moving pulsar does exist in the G327.1-1.1, it will be interesting to examine its relationship to the Cos-B source CG327-0. However, CG327-0 has
not been confirmed in the recent and more sensitive 100 MeV gamma-ray catalogs of \egret\ . Further
investigation may be needed.

\section{CONCLUSION}
The morphological and spectral analyses of the \asca\ , \rosat\ and MOST radio data reveal that
G327.1-1.1 is obviously a composite SNR. The X-ray synchrotron
nebula may be powered by an undiscovered pulsar which may now be located near the X-ray common peak, or 
the head of the radio ridge.
 The X-ray spectrum of G327.1-1.1 is well fitted by two components, a power-law component
with photon index of $\sim$ 2.2 and a thermal component with temperature $\sim$ 0.4keV. From the best fit of the thermal component, the explosion energy of the supernova, the ISM density, the shock 
velocity and the age
of G327.1-1.1 are derived as shown in Table 2. From the best fit of the nonthermal component, 
we predict the period, the period derivative of the pulsar, and the strength of its surface magnetic 
field as shown in $\S$4.3. These quantities are all in the range of reasonable values for a typical
pulsar. The magnetic field of the synchrotron nebula is also estimated. The positional
coincidence between the unusual X-ray elongation and the radio ridge may imply the existence
of a trailing configuration produced by a run-away pulsar at a velocity of $>$ $600{\rm km s^{-1}}$. 
The emission from the ridge (or the trail) is ascribed to arise from the synchrotron nebula confined
by a bow shock. 
We expect that, with the development of the
X-ray satellite, more and more composite SNRs will be found.  
And for this source, we wish further observations would be carried out by \xmm\ or \axaf\ to provide more detailed
information.

We would like to thank Jiong Qiu for kind help in overlaying the images and thank Bryan Gaensler for
providing the MOST data in electronic form. Sun would like to thank Xun Zhang for kind help in
english. This research has made use of data obtained through the High Energy
Astrophysics Science Archive Research Center (HEASARC), provided by NASA's Goddard Space Flight Center. This work is carried out in the Lab of Astronomical Data Analysis in Nanjing University. It is supported
by the Ascending Project of the State Scientific Committee of China, the National Natural Science
Foundation of China, and the Foundations for training PhD students and supporting repatriating
scholars from the State Education Division of China.

\clearpage
\begin{figure}
\vspace{-1.0in}
\figcaption{\asca\ SIS spectra of G327.1-1.1 . {\it Upper panel}: Raw spectrum with best-fit
power-law model plus thermal thin plasma model. {\it Lower panel}: Residuals between data and best-fit model.}
\end{figure}

\begin{figure}
\vspace{-1.0in}
\figcaption{Best-fit model. Thermal thin plasma model is in dash-dot line and power-law model is in dash line. Their sum is in solid line. 
The  power-law component is dominant, while the thermal component is only distinct below 2 keV.}
\end{figure}

\begin{figure}
\vspace{-1.0in}
\figcaption{The SIS soft X-ray(0.4-2 keV) deconvoluted image is shown in dashed contours and the 
SIS hard X-ray(2-10 keV) deconvoluted image is shown in solid contours, while the \rosat\ PSPC 
image is shown
in gray-scale. The two SIS images are all corrected for exposure, vignetting and background. The contour levels are 0.2, 0.35, 0.5, 0.65, 0.8, 0.95 of the maximum.
 PSPC image is smoothed with a Gaussian function of ${\sigma}$=$30\myarcsec$. 
 The unresolved source 4 has been removed to show the SNR emission more clearly.}
\end{figure}

\begin{figure}
\vspace{-1.0in}
\figcaption{The \rosat\ grey image(smoothed with a Gaussian function with FWHM=30$\myarcsec$ ) overlay the radio contour map in 843MHz.
 The two brightest source in \rosat\ image were attributed to foreground or
background sources by Seward, et al.(1996). The radio shell is nearly complete except in the northeast
and northwest. The radio plerionic structure is 
off-center and has two peaks. The radio long ridge is well in positional coincidence with the main part of the \rosat\ X-ray emission. We add some levels into the original linear scale levels to show the
plerionic
structure and the shell more clearly, so radio contour levels are in 0.02, 0.05, 0.15, 0.3, 0.45, 0.6, 0.75, 0.9,
0.95 of the maximum.} 
\end{figure}
\clearpage

\newpage
\footnotesize
\centerline{\bf Table 1}
\vspace{0.5cm}
\centerline{\bf Spectral fitting result of a single Power Law , or Power Law plus Thermal Thin Plasma Model$^{\rm\  a}$}
\vspace{0.5cm}
\doublerulesep 1pt
 \centerline{\begin{tabular}{c|c|c}\hline\hline
   Parameter   &    Value$^{\rm\  c}$(only nonthermal)  &  Value$^{\rm\  c}$(nonthermal and thermal)\\\hline
   $N_{\rm H}(10^{22}$cm$^{-2})$  &  1.14$^{+0.12}_{-0.10}$   &    1.8$^{+0.3}_{-0.3}$ \\
   Photon Index    &   2.01$^{+0.12}_{-0.10}$  &   2.2$^{+0.2}_{-0.2}$ \\
   {\it kT}(keV)   &   -    &  0.37$^{+0.35}_{-0.20}$ \\
   $F_{\rm x}^{\rm\ b}$(observed nonthermal) &  5.9$\times10^{-12}$ergs$\mbox{ }$cm$^{-2}$s$^{-1}$  &  5.4$\times10^{-12}$ergs$\mbox{ }$cm$^{-2}$s$^{-1}$  \\
   F$_{\rm x}^{\rm\ b}$(unabsorded nonthermal) & 1.1$\times10^{-11}$ergs$\mbox{ }$cm$^{-2}$s$^{-1}$ & 1.2$\times10^{-11}$ergs$\mbox{ }$cm$^{-2}$s$^{-1}$ \\ 
   F$_{\rm x}^{\rm\ b}$(observed thermal)   &   -   &   3.5$\times10^{-13}$ergs$\mbox{ }$cm$^{-2}$s$^{-1}$\\
   F$_{\rm x}^{\rm\ b}$(unabsorded thermal) &   -  &  2.8$\times10^{-11}$ergs$\mbox{ }$cm$^{-2}$s$^{-1}$\\
   $\chi^{2}$    &   262.1/277(d.o.f)  &   240.6/275(d.o.f)\\ \hline\hline
 \end{tabular}}
\begin{flushleft}
\leftskip 18pt
$^{\rm a}$ Cosmic abundances (Anders \& Grevesse 1989)\\
$^{\rm b}$ F$_{\rm x}$ is the flux in 0.5keV - 10keV band.\\
$^{\rm c}$ Single-parameter 2.706~$\sigma$ errors(90$\%$ confidence region)\\ 
Notes: We used data of SIS0, SIS1 and PSPC to do a joint fit.
\end{flushleft}

\newpage
\centerline{\bf Table 2}
\vspace{0.5cm}
\centerline{\bf Properties about the thermal component}
\vspace{0.5cm}
\doublerulesep 1pt
 \centerline{\begin{tabular}{c|c}\hline\hline
   Property     &    Value$^{\rm\ a\mbox{ }b}$ \\\hline
   Fitted Properties$^{\rm\ c}$:  &        \\
   $N_{\rm H}(10^{22}$cm$^{-2})$  & (1.8) \\
   {\it kT}(keV)  &  0.40$^{+0.34}_{-0.15}$ \\
   $F_{\rm x}$(0.5-2keV)  &   7.5$\times10^{-13}$ergs$\mbox{ }$cm$^{-2}$s$^{-1}$\\
   Unabsorbed $F_{\rm x}$(0.5-2keV)  &  6.2$\times10^{-11}$ergs$\mbox{ }$cm$^{-2}$s$^{-1}$\\
   Derived Properties$^{\rm\ d}$:  &        \\
   D(kpc)     &    (9.0)\\
   R(pc)    &   17.0\\
   n$_{0}($cm$^{-3})$  &   0.10\\
   t$_{4}(10^{4}$yr)   &  1.1\\
   M$_{\rm su}$(M$_{\odot})$  &  49\\
   E$_{0}(10^{51} $ergs)  &  0.23\\
   $\upsilon_{\rm s}$(km s$^{-1}$)  & 600 \\ \hline\hline
 \end{tabular}}
\begin{flushleft}
\leftskip 90pt
$^{\rm a}$ Values in bracket are fixed.\\
$^{\rm b}$ Single-parameter 2.706~$\sigma$ errors(90$\%$ confidence region)\\
$^{\rm c}$ These properties are the best fit of \rosat\ data to mainly outer\\ 
\leftskip 100pt
           part of G327.1-1.1.\\
\leftskip 90pt
$^{\rm d}$ We synthesize the fitting results from \asca\ data and \rosat\ \\
\leftskip 100pt
           to calculate these properties.\\
\end{flushleft}

\newpage
\pagestyle{empty}
\normalsize
\vspace{-1.0in}
\begin{figure}[p]
\vspace{-1.0in}
\centerline{\psfig{file=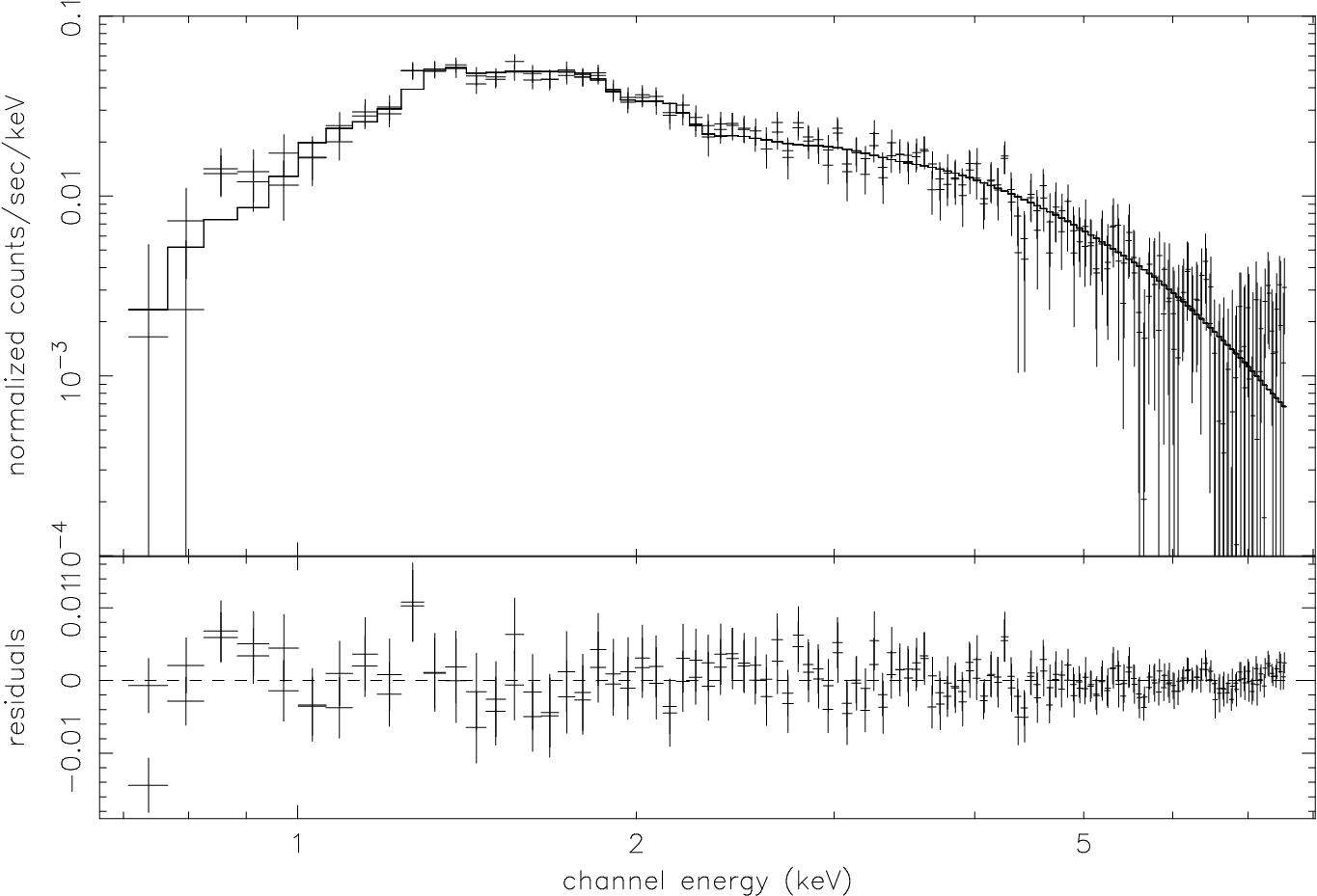,width=5in,angle=270}}
\end{figure}

\begin{figure}[t]
\vspace{-2.0in}
\centerline{\psfig{file=fig2.ps,width=5in,angle=270}}
\end{figure}
\clearpage
 
\newpage
\begin{figure}[t]
\centerline{\psfig{file=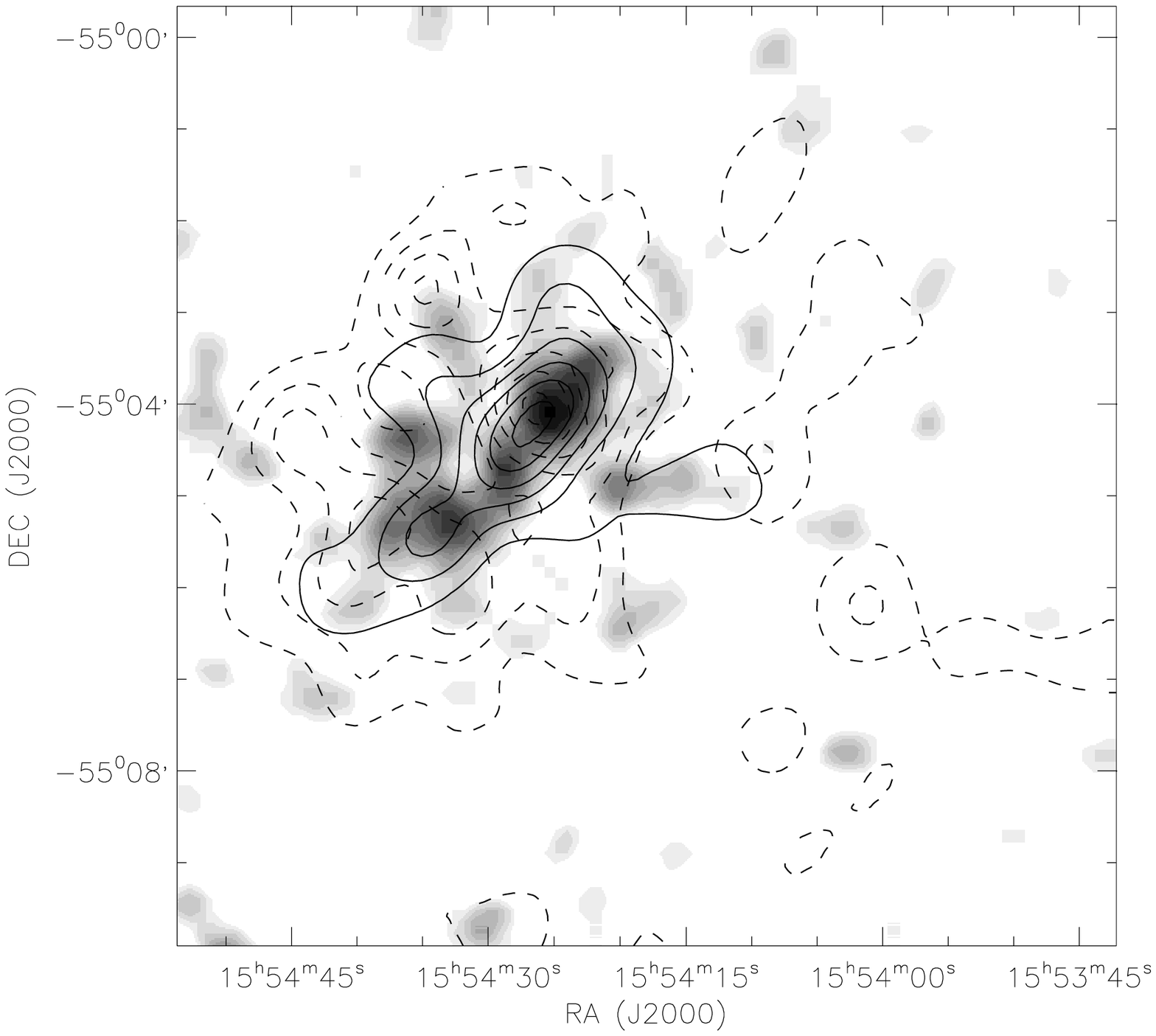,width=6in}}
\end{figure}
 
\newpage
\begin{figure}[t]
\centerline{\psfig{file=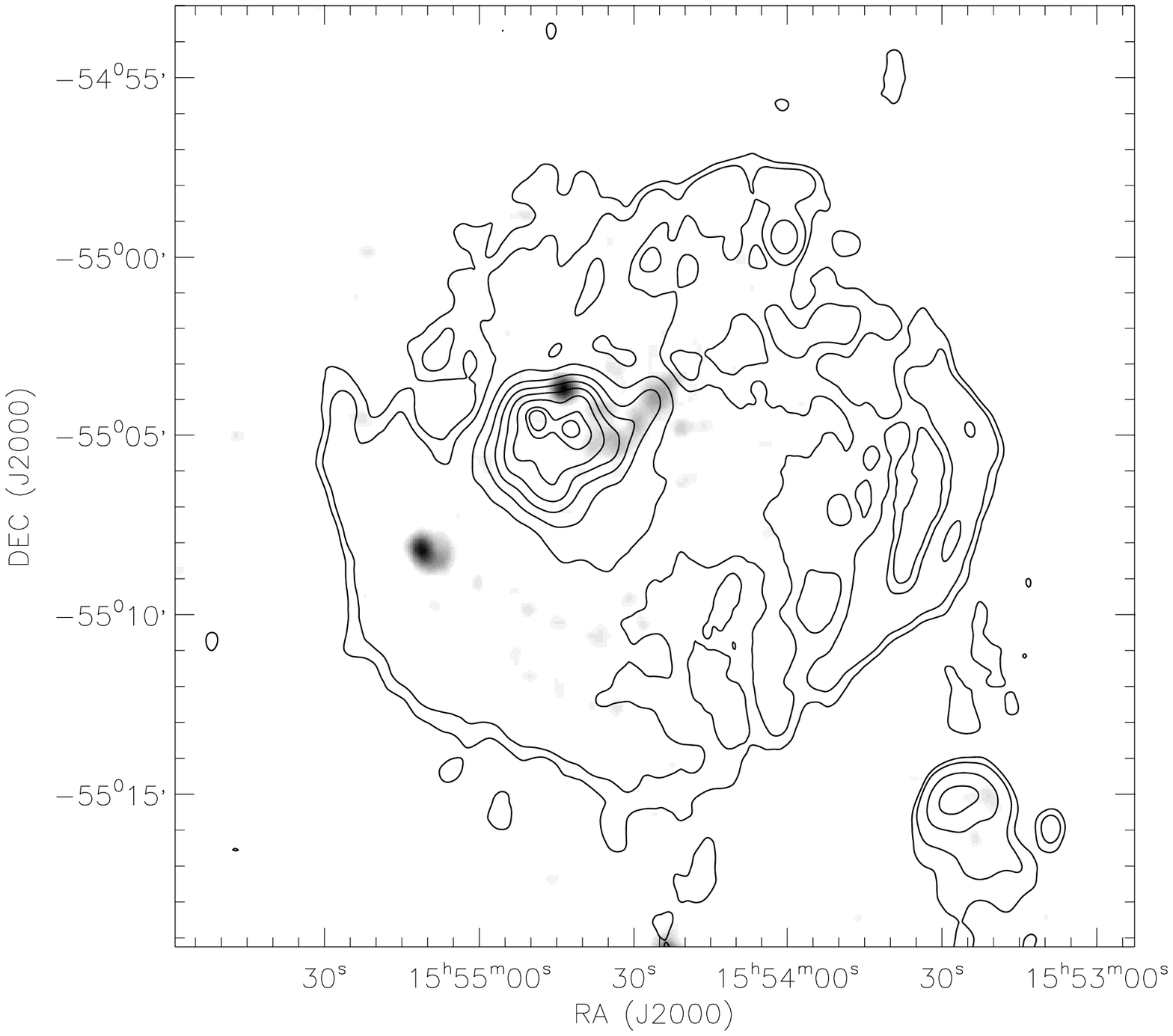,width=6in}}
\end{figure}

\end{document}